    \documentclass{JHEP3}
    \usepackage{amsmath}
    \usepackage{amssymb}
    \usepackage{graphicx}
    \numberwithin{equation}{section}
    \usepackage{epsfig}

    \renewcommand{\[}{\left[}
    \renewcommand{\]}{\right]}
    \renewcommand{\(}{\left(}
    \renewcommand{\)}{\right)}

    \newcommand{\ph}{{\rm ph}}
    \newcommand{\mir}{{\rm mir}}

    \newcommand{\beq}{\begin{equation}}
    \newcommand{\eeq}{\end{equation}}
    \newcommand\beqa{\begin{eqnarray}}
    \newcommand\eeqa{\end{eqnarray}}
    \newcommand\bea{\begin{array}}
    \newcommand\eea{\end{array}}

    \newcommand\RE{{\rm Re}\,}

    \def\XXint#1#2#3{{\setbox0=\hbox{$#1{#2#3}{\int}$}
    \vcenter{\hbox{$#2#3$}}\kern-.5\wd0}}

    \newcommand{\nn}{\nonumber}

    \newcommand{\COMMENT}[1]{}
    
    \newcommand{\neqa}{\nonumber\end{eqnarray}}
    \newcommand{\la}[1]{\label{#1}}

    \newcommand{\eq}[1]{(\ref{#1})}

    \newcommand{\half}{\frac{1}{2}}

    \renewcommand{\d}{\partial}
    
    \newcommand{\<}{{\langle}}
    \renewcommand{\>}{{\rangle}}

    \newcommand{\re}{\relax{\rm I\kern-.18em R}}

    \newcommand{\rb}{\right)}
    \newcommand{\lb}{\left(}

    \def\su2{{SU(2)}}

    \def\<{\langle}
    \def\>{\rangle}

    \def\i2{\frac{i}{2}}

    \newcommand{\be}{\begin{equation}}
    \newcommand{\ee}{\end{equation}}
    \newcommand{\bg}{\begin{gather}}
    \newcommand{\eg}{\end{gather}}
    \newcommand{\bseq}{\begin{subequations}}
    \newcommand{\eseq}{\end{subequations}}
    \def\half{\frac{1}{2}}

    \newcommand{\bac}{\begin{array}{l}}
    \newcommand{\eac}{\end{array}}

    \newcommand{\bal}{\begin{array}{l}}
    \newcommand{\eal}{\end{array}}

\newcommand{\AdSSfive}{{\rm AdS}_5\times {\rm S}^5}

    \title{Y-system and $\beta$-deformed N=4 Super--Yang-Mills}
    \author{
    Nikolay Gromov\\ Mathematics Department, King's College London,
      The Strand, London WC2R 2LS, UK \& St.Petersburg INP, Gatchina, 188 300, St.Petersburg, Russia\\
    E-mail: \email{nikolay.gromov$\bullet$kcl.ac.uk}}
    \author{Fedor~Levkovich-Maslyuk\\ Physics Department, Moscow State University, 119991, Moscow, Russia\\
    E-mail: \email{fedor.levkovich$\bullet$gmail.com}}

    \abstract{We show how the perturbation theory results recently obtained by F.Fiamberti,
    A.Santambrogio, C.Sieg and D.Zanon for operator anomalous dimensions of $\beta$-deformed Super-Yang-Mills theory
    can be reproduced from the AdS$_5$/CFT$_4$ Y-system proposed by N.G., V.Kazakov and P.Vieira.
    To do this, we obtain the general twisted asymptotic solution of this Y-system of functional equations.
    We show that existence of an additional parameter $\beta$ in the deformed theory allows
    to extract rich information about the perturbation theory integrals directly from Y-system. Using this method
    we found a simple generating function for a broad class of such integrals.

    }

    \preprint{}
    
\begin{document}

\section{Introduction}

The celebrated AdS/CFT correspondence relates a gauge field theory and a string theory, with the best-studied
 example being the duality between four-dimensional ${\cal N}=4$ planar
    superconformal Yang-Mills (SYM) theory and Type IIB superstring theory on
    $\AdSSfive$ \cite{AdS/CFT}. Recently more similar
    examples of dualities where found \cite{AdSmore,zarads3}.  Integrability properties, which have been discovered on both sides of such dualities, have played an important role in the study of this rapidly
    developing subject. The exact S-matrix led to formulation of asymptotic Bethe ansatz equations (ABA)
    \cite{adsbae,baedual,Crossing,zarads3}, which describe the anomalous dimensions for operators of asymptotically large length $L$ at any coupling.
    The generalized L\"uscher formula \cite{luscherref}, Y-system \cite{GKV} and Thermodynamic Bethe Ansatz
    \cite{tbarefsads5,tbarefscp3,GFcp3} have made it possible to
    take into account the wrapping corrections and obtain, in principle, the missing part of the spectrum at finite $L$.

In the 4D case, evidence for integrability has been found also for the $\beta$-deformed SYM theory, which has ${\cal N}=1$ instead of ${\cal N}=4$ supersymmetry. The deformation consists in replacing the original superpotential for the chiral superfields by

\beq
W= ih\ {\rm tr}
(e^{i\pi\beta}\phi\psi Z-e^{-i\pi\beta}\phi Z\psi).
\eeq

The deformed theory remains superconformal in the planar limit to all orders of perturbation theory
\cite{LeiStr,MPSZ} if $\beta$
is real and $h{\overline h}=g_{\rm YM}^2$,
where $g_{\rm YM}$ is the Yang-Mills coupling constant, related to the 't Hooft coupling $g^2$ in the planar limit as
\beq
g^2=\frac{g_{\rm YM}^2 N}{16\pi^2}\;.
\eeq
Under these conditions the deformation becomes exactly marginal. The $\beta$-deformed theory is also believed to have a string dual \cite{LunMal}. Integrabibility properties of that string theory have been studied in \cite{Frolov,BykFro}.

The deformed theory was also investigated quite intensively in the perturbative regime. Evidence for perturbative integrability was found in \cite{FrRoTs, Frolov:2005iq,BerChe,BeiRoi,Roiban:2003dw}. On the other hand, direct computations of anomalous dimensions without use of integrability were done in \cite{FSSZ,FSSZ1} (see also \cite{Fiamberti:2010fw}). In those works, wrapping corrections at critical order have been found for two operators of length $L=4$ with two impurities, and for one-impurity operators with $L \leq 11$. Recurrence relations were also discussed \cite{FSSZ1, Fiamberti:2010fw} which allow one in principle to obtain this correction for any one-impurity operator, though a closed formula for the corrections was not found.

The methods which rely on integrability have reproduced only a part of the results.
In \cite{Gunnesson:2009nn,Beccaria:2009hg} first wrapping corrections were obtained for certain single impurity operators, though only for $\beta=\half$. Also, very recently a part of the S-matrix was presented as a conjecture, and made it possible to reproduce the first wrapping correction to the $L=4$ Konishi operator via generalised L\"uscher formula \cite{Ahn:2010yv}
which gave strong support for the integrability for arbitrary real values of $\beta$.

For ${\cal N}=4$ SYM another efficient method, based on the asymptotic large $L$ solution of the Y-system \cite{GKV},
was used in \cite{GKV, Fiamberti:2009jw} to analytically compute wrapping corrections,
giving perfect agreement with direct perturbative results \cite{Fiamberti:2008sh,Fiamberti:2009jw}.
At the leading wrapping order the Y-system should be equivalent to the generalized
L\"usher formula of \cite{luscherref}.
 Here we argue that the Y-system of \cite{GKV} describes also the $\beta$-deformed theory, and present a generalised version of that asymptotic solution with $4$ additional twist parameters. We show that it reproduces
all perturbative results of \cite{FSSZ,FSSZ1} for $\beta$-deformed SYM. In particular we study the one magnon case
in detail, giving a general formula for the first wrapping correction
for a single impurity operator of arbitrary length $L$.

\section{The asymptotic solution of Y-system}
In this section we briefly describe the general Y-system technique
and the generating functional which allows to build the
asymptotic large $L$ solution of the Y-system and T-system of \cite{GKV}.
We then propose a way to modify this functional for the $\beta$-deformed theory.

\subsection{Review of Y- and T-systems}

\FIGURE[ht]
{\label{Fig:figysys}

    \begin{tabular}{cc}
    \includegraphics[scale=0.6]{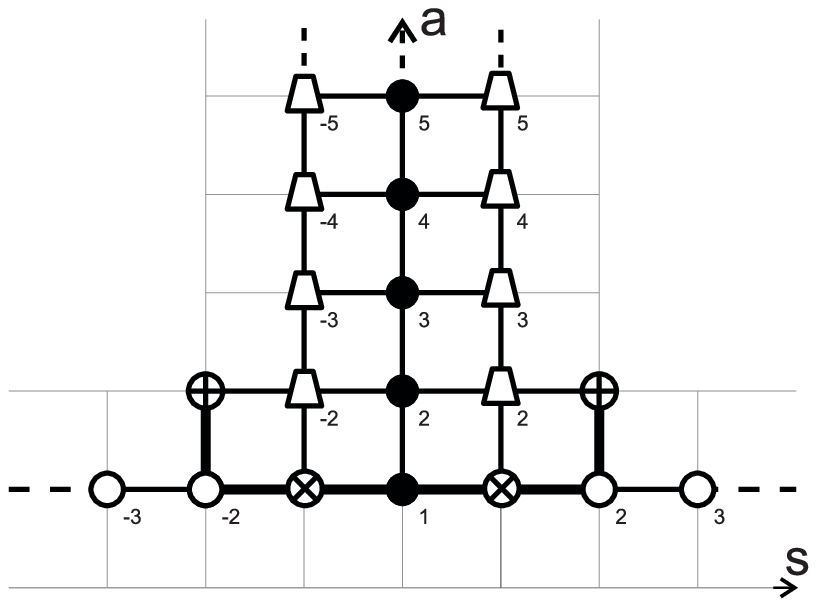}
    \\
    Y-system and
    T-system
    \end{tabular}
    \caption{Graphical representation of the Y-system and T-system \cite{GKV}.
    Nodes marked by
    small symbols
    correspond to Y-functions. Intersections of grid lines in the T-hook correspond to T-functions.}
}

The Metsaev-Tseytlin $\AdSSfive$ string action in the light-cone gauge is a classically integrable
2D field theory, and its energy spectrum is believed to describe the spectrum of anomalous dimensions of planar ${\cal N}=4$ SYM.
In general, the experience with relativistic integrable theories \cite{Zamolodchikov:1991et} suggests that
the exact quantum spectrum should be governed by a system of functional
Hirota equations\footnote{sometimes they could be slightly more complicated}
\beq
\label{Tsystem}
  T_{a,s}(u+i/2)T_{a,s}(u-i/2) =T_{a+1,s}(u)T_{a-1,s}(u)+T_{a,s+1}(u)T_{a,s-1}(u)\;.
\eeq
We use here the following short-hand notations
    \be
        f^\pm \equiv f(u\pm i/2), \ \ \ f^{[+a]} \equiv f(u + ia/2).
    \ee
It was conjectured in \cite{GKV} that in the AdS/CFT case the system of Hirota equations
should be exactly the same, with the functions $T_{a,s}(u)$ being non-zero
only inside the infinite T-shaped domain of the ${a,s}$ integer lattice, shown in Fig.\ref{Fig:figysys}.

In order to compute physical quantities one should form particular
combinations of these $T$-functions
\begin{equation}
\label{YT}
Y_{a,s}=\frac{T_{a,s+1}T_{a,s-1}}{T_{a+1,s}T_{a-1,s}}\;\; .
\end{equation}
As a consequence of \eq{Tsystem}, the Y-functions satisfy the Y-system functional equations\footnote{The equations for $\{ a,s\} =\{2, 2\}$ and $\{a,s\} =\{-2, 2\}$ cannot be written in such ``local'' form.}
\begin{equation}
\label{eq:Ysystem}
\frac{Y_{a,s}^+ Y_{a,s}^-}{Y_{a+1,s}Y_{a-1,s}}
  =\frac{(1+Y_{a,s+1})(1+Y_{a,s-1})}{(1+Y_{a+1,s})(1+Y_{a-1,s})} \,.
\end{equation}
Indices ${a,s}$ here label the marked nodes of the lattice in Fig.\ref{Fig:figysys}.

The Y-system should be supplemented with a particular set of analytical properties.
One possibility was proposed in \cite{Cavaglia:2010nm}.
In the current case the analytical properties are rather involved,
partly due to the lack of Lorentz symmetry and partly due to the
complicated $psu(2,2|4)$ symmetry of the theory.
In particular the dispersion relation for a single excitation in infinite volume is quite nontrivial.

We express the energy and momentum of the excitations (also called magnons)
in terms of the
Zhukowski variable $x(u)$, defined by

    \be
    \label{xdef}
        x + \frac{1}{x} = \frac{u}{g}.
    \ee
The ``mirror" and ``physical" branches of this function are defined as
   \beq
    \label{xBranches}
        x^{\ph}(u)=\frac{1}{2}\lb
        \frac{u}{g}+\sqrt{\frac{u}{g}-2}\;\sqrt{\frac{u}{g}+2}\rb\;\;,\;\;
        x^{\mir}(u)=\frac{1}{2}\lb \frac{u}{g}+i\sqrt{4-\frac{u^2}{g^2}}\rb \,,
    \eeq
    where $\sqrt{u}$ denotes the principal branch of the square root. The energy and momentum of a bound state with $n$
    magnons are given by
 \begin{equation}
\label{epsandp}
\epsilon_n(u)= n+\frac{2ig}{x^{[+n]}}-\frac{2ig}{x^{[-n]}}\;\;,\;\;
p_n(u)=\frac{1}{i}\log\frac{x^{[+n]}}{x^{[-n]}}\;.
\end{equation}
Finally, the exact energy of a state is given by the expression
\beq
\label{Efull}
E=\sum_{j}\epsilon_1^\ph (u_{4,j})+\delta E\;\;,\;\;
\delta E = \sum_{a=1}^\infty \int\frac{du}{2\pi i}
\,\,\frac{\d\epsilon_a^\mir (u)}{\d u} \log\left(1+Y_{a,0}^\mir (u)\right)\;,
\eeq
with the rapidities $u_{4,j}$ being fixed by the exact Bethe ansatz equations
\beq
Y_{1,0}^\ph (u_{4,j})=-1\,.
\eeq
where $Y_{1,0}^\ph(u)$, similarly to $x^\ph$, is the result of the analytical
continuation of $Y_{1,0}^\mir(u)$ through the cut $(i/2+2g, i/2+\infty)$ \cite{GKVKonishi}.

The Y-system for ${\cal N}=4$ SYM passes several nontrivial tests: it reproduces both perturbative wrapping corrections \cite{GKV,Fiamberti:2009jw} and quasiclassical spectrum at strong coupling \cite{ads5quasicl}, and is moreover compatible with Thermodynamic Bethe ansatz
equations \cite{tbarefsads5} (which allow also efficient numerical studies \cite{GKVKonishi,tbanum}).

In this paper we give evidence that exactly the same Y-system set of equations describes the
$\beta$-deformed theory. We show that the asymptotic solution of \cite{GKV}
is in fact a representative of a family of solutions, which have similar analytical properties
and in terms of the transfer matrices correspond to the twisted case\footnote{Usually the
construction of transfer matrices allows to introduce extra twist parameters
without destroying integrability. Often the twisted systems
can be better controlled and in some cases the twists are necessary as
regularizations, see for example \cite{Bazhanov:2010ts,GV1loop}.}.

\subsection{Twisted generating functional}

%In \cite{Kazakov:2007fy,Zabrodin:2007rq} an efficient method of constructing solutions of the Hirota equation
%for a domain called ${\mathbb L}$-hook (one half of the ${\mathbb T}$-hook diagram in Fig.1) was described in detail. It is based on the Backlund transformation
%which allows to gradually reduce the domain to a trivial one.
%This method can be also applied for the ${\mathbb T}$-hook.

In \cite{Tsuboi:1997iq,Tsuboi:1998ne,Kazakov:2007fy, Zabrodin:2007rq} a method was proposed for constructing solutions of the Hirota
equation for a domain called ${\mathbb L}$-hook (one half of the ${\mathbb T}$-hook diagram in Fig.1).
The method relies on the use of Wronskian relations and Backlund transformations which allow to gradually reduce the domain to a trivial one.
The result obtained in this way can be written compactly in terms of a generating functional.
A similar result was recently obtained for the ${\mathbb T}$-hook \cite{Zengo}.
%The generating functional for the T-functions in this case can be interpreted as the generating
%functional for transfer matrices with $u(2,2|4)$ symmetry.

In this paper we want to demonstrate that the twisted solution
of the Hirota equation (see \cite{Zabrodin:2007rq}) can indeed be used for the $\beta$-deformed theory.
For that we just need to find the asymptotic large $L$ solution,
which can then be applied also for comparison with perturbation theory up to
order $\sim g^{4L-2}$. In the large $L$ limit the ``massive" nodes $Y_{a,0}$ are suppressed
and the Y-system decouples into two wings: $su_{\rm L}(2|2)$ and $su_{\rm R}(2|2)$. The solution
for a single wing is much simpler than for the full $psu(2,2|4)$ case, and it is given by an explicitly known generating functional. Here we propose to use the following twisted version of that functional\footnote{Here we use $su(2)$ grading.
For this grading the method of \cite{GKV} is described in detail in \cite{Serban:2010sr}.}$^,$\footnote{N.G. thanks P.Vieira for the discussion of this possibility and for the collaboration on the
early stages of this work.}:
\beqa
\label{genWR}
    {\cal W}_{\rm R}&=&
    \frac{1}{1-\frac{1}{\tau_{1,\rm R}}D\frac{B^{(-)-}Q_1^+}{B^{(+)-}Q_1^-}D}
    {\(1-\frac{1}{\tau_{2,\rm R}}D\frac{Q_1^+Q_2^{--}}{Q_1^-Q_2}D\)}\\
    &\times&\nn
    {\(1-\tau_{2,\rm R}D\frac{Q_2^{++}Q_3^-}{Q_2^{}Q_3^+}D\)}
    \frac{1}{1-\tau_{1,\rm R}D\frac{R^{(+)+}Q_3^-}{R^{(-)+}Q_3^+}D}\;,
\eeqa
where $\tau_{1,\rm R}, \; \tau_{2,\rm R}$ are complex numbers (not dependent on the spectral parameter $u$) which we call twists.
The generating functional for the left wing ${\cal W}_{\rm L}$ is given by the same expression with
$Q_{1,2,3}$ replaced by $Q_{7,6,5}$ and $\tau_{1,2\; \rm R}$ by $\tau_{1,2\; \rm L}$.
We use the following notation:
\beq
Q_l \equiv \prod_{j=1}^{K_l} (u-u_{l,j})\;\;,\;\;
R^{(\pm)}\equiv\prod_{j=1}^{K_l}\(x(u)-x^{\ph\mp}_{4,j}\)\;\;,\;\;
\eeq
\beq
B^{(\pm)}_l\equiv\prod_{j=1}^{K_l}\(\frac{1}{x(u)}-x^{\ph \mp}_{l,j}\), \ B^{(\pm)} \equiv B^{(\pm)}_4\;,
\eeq
and $D=e^{-i\d_u/2}$ is the shift operator.
Expansion of this functional gives the functions $T_{a,1}^{{\rm R},{\rm L}}$ and $T_{1,s}^{{\rm R},{\rm L}}$:
\beq
\label{WTsa}
{{\cal W}}_{{\rm R},{\rm L}}=\sum_{s=0}^{\infty}D^s T_{1,s}^{{{\rm R},{\rm L}}}D^s
\;\;,\;\;
{\cal W}^{-1}_{{\rm R},{\rm L}}=\sum_{a=0}^{\infty}(-1)^aD^a T_{a,1}^{{\rm R},{\rm L}}D^a\;.
\eeq
Let us motivate the structure of the twists we introduced above.
One could introduce eight twists in total: one in each of the four terms inside ${\cal W_{{\rm R}}}$ and, similarly, four more twists in ${\cal W_{{\rm L}}}$. However,
it is easy to see that requiring $Y_{1,s}$ and $Y_{a,1}$ to be real
implies that the twists in the first and last terms inside ${\cal W}$ are complex conjugate to each other, and the same is true for twists in
the second and third terms.
Also, to allow only such configurations of Bethe roots which are
invariant w.r.t. complex conjugation one should require the twists to be unimodular.
Thus the generating functional satisfying these requirements could have only
four independent twists in total: two in ${\cal W_{{\rm L}}}$ and two in ${\cal W_{{\rm R}}}$, as it is indeed written in \eq{genWR}.

The polynomials $Q_a(u)$ in the denominators of generating functionals could potentially
result in the poles of the $T_{a,s}$ functions. However, one can show that these poles
cancel provided the following Bethe equations are satisfied\footnote{with the $x^\ph$ branch used for $x(u)$ in all terms}
\beq
1=\left.\frac{\tau_{2,\rm R}}{\tau_{1,\rm R}}\frac{B^{(-)}}{B^{(+)}}
\frac{Q^+_2}{Q_2^{-}}\right|_{u=u_{1,k}}
,\;\;
-1=\left.\frac{1}{(\tau_{2,\rm R})^2}\frac{Q_1^+Q_2^{--}Q_3^+}{Q_1^-Q_2^{++}Q_3^-}
\right|_{u=u_{2,k}}
,\;\;
1=\left.\frac{\tau_{2,\rm R}}{\tau_{1,\rm R}}\frac{R^{(-)}}{R^{(+)}}
\frac{Q^+_2}{Q_2^{-}}\right|_{u=u_{3,k}}\la{ABA}
\eeq
with a similar set of $3$ equations for the left wing.

For large $L$, the middle node $Y_{a,0}$ is given by \cite{GKV}
\beq
\label{Yaasymp}
Y_{a,0}\simeq T_{a,1}^{\rm L} T_{a,1}^{\rm R}
\prod^{\frac{a-1}{2}}_{n=-\frac{a-1}{2}}\Phi(u+in)\;,
\eeq
which can be found by solving \eq{eq:Ysystem}
for $s=0$. Here $\Phi$ is the only unknown function which is almost fixed by the
requirements that $Y_{a,0}$
is real and that $Y_{a,0}^\ph(u_{4,j})$
is unimodular as a function of $u_{4,j}$. Those conditions are satisfied
as a consequence of the crossing equation \cite{Crossing, GKV} by the following expression\footnote{Here we use the gauge of  \cite{Serban:2010sr} which is different from the one in \cite{GKV}.}
\beq
\label{defPhi}
\;\;\Phi(u)=\(\frac{x^-}{x^+}\)^{ L}\prod_{j=1}^{K_4}\sigma^2(u,u_{4,j})
\(\frac{R^{(-)+}}{R^{(+)+}}\)^2\frac{Q_4^{++}}{Q_4^{--}}\frac{B_1^-B_3^+B_5^+B_7^-}{B_1^+B_3^-B_5^-B_7^+}\;.
\eeq
The equation for the momentum-carrying roots $Y_{1,0}^\ph (u_{4,k})=-1$ reads
\beq
\label{bae4fromW}
\left.
\tau_{1,\rm R}\tau_{1,\rm L}\Phi(u)\(\frac{R^{(+)+}}{R^{(-)+}}\)^2\frac{Q_3^-Q_5^-}{Q_3^+Q_5^+}
\right|_{u=u_{4,k}}
=-1\;.
\eeq
It's important to mention that the Bethe equations \eq{ABA}, \eq{bae4fromW} are consistent with the ABA of \cite{BeiRoi}. For more details on this see Appendix A, in which we also describe the switch to $sl(2)$ grading.

In the next section we consider restriction to the $su(2)$ subsector
and study the weak coupling limit of these expressions.

\section{$su(2)$ subsector}\la{sec:su2}
For the $su(2)$ subsector only $u_{4,j}$ roots are introduced, and the Bethe ansatz equations read \cite{FrRoTs,BeiRoi,Roiban:2003dw,BerChe}
\beq
\label{bae4}
\(\frac{x^{+}_{4,k}}{x^-_{4,k}}\)^L = q^{2L}\prod^{K_4}_{j\neq k}\sigma^2(u_k,u_j)\frac{u_{4,k}-u_{4,j}+i}{u_{4,k}-u_{4,j}-i}
\;,
\eeq
where $q\equiv\exp(\pi i\beta)$.
For this equation to coincide with \eq{bae4fromW}, the equality $\tau_{1,\rm L}\tau_{1,\rm R} = q^{2L}$ must hold.
Furthermore, we found that in order to match our explicit answers for anomalous dimensions with the many perturbative results
 we have to set
\beq
\label{twistsu2}
\tau_{1,\rm L} = q^{2L-2K_4}\;\;,\;\;
\tau_{1,\rm R} = q^{2K_4}\;\;,\;\;
\tau_{2,\rm R} = \tau_{2,\rm L} = 1\;\;.
\eeq
These expressions are also in agreement with the values of twists $\tau$ obtained by comparing our ABA equations \eq{ABA}, \eq{bae4fromW} with the ABA equations obtained in \cite{BeiRoi} for the $\beta$-deformed theory (see Appendix A), which gives additional support for the ABA of \cite{BeiRoi}.

For the $su(2)$ subsector $Q_a=1$ if $a\neq4$, and we get an explicit expression from the generating functional:
\beq
\label{Ta}
(-1)^aT^{{\rm R}}_{a,1} =(a+1)-a \tau_{1,{\rm R}}\frac{R^{(+)[+a]}}{R^{(-)[+a]}}
-a \frac{1}{\tau_{1,{\rm R}}}\frac{B^{(-)[-a]}}{B^{(+)[-a]}}
+(a-1)\frac{R^{(+)[+a]}}{R^{(-)[+a]}}\frac{B^{(-)[-a]}}{B^{(+)[-a]}}\;.
\eeq
We see that indeed $T^{{\rm R}}_{a,1}$ and $T^{{\rm L}}_{a,1}$ are real functions for all $a$, since $\overline{ R^{(\pm)}}=B^{(\mp)}$,
and hence $Y_{a,0}$ is also real. $Y_{a,0}$ is given by \eq{Yaasymp} with $\Phi$ obtained from \eq{defPhi}:
\beq
\;\;\Phi(u)=\(\frac{x^-}{x^+}\)^{L}\prod_{j=1}^{K_4}\sigma^2(u,u_{4,j})
\(\frac{R^{(-)+}}{R^{(+)+}}\)^2\frac{Q_4^{++}}{Q_4^{--}}\;.
\eeq

\section{Weak coupling expansion}\la{sec:weak}
To obtain the leading wrapping correction to operator anomalous dimensions, we insert into \eq{Efull} the Y-functions given by \eq{Yaasymp} and expand them at
weak coupling, as in \cite{GKV,Serban:2010sr}.
For $g\to0$ we have
\beq
\label{weak1}
\frac{R^{(+)[+a]}}{R^{(-)[+a]}}\simeq\frac{Q_4^{[+a+1]}}{Q_4^{[+a-1]}}\;\;,\;\;
\frac{B^{(-)[-a]}}{B^{(+)[-a]}}\simeq\frac{Q_4^{[-a-1]}}{Q_4^{[-a+1]}}\;,
\eeq
\beq
\label{weak2}
\frac{x^{\mir [-a]}}{x^{\mir [+a]}} \simeq \frac{4g^2}{a^2+4u^2}\;\;,\;\;
\Phi_a\simeq\(\frac{4g^2}{a^2+4u^2}\)^L\frac{Q_4^{[+a-1]}Q_4^{[-a+1]}}{Q_4^{[+a+1]}Q_4^{[-a-1]}}\;,
\eeq
and hence $T_{a,1}^{\rm R,L}$ are rational functions of $u$. In addition,
\beq
\frac{\d\epsilon_a^\mir (u)}{\d u} \simeq -2i\;,
\eeq
so that \eq{Efull} can be written as
\beq\la{integ}
\delta E \simeq -\sum_{a=1}^{\infty}\int\limits \frac{d u }{\pi}
Y_{a,0}(u)
.
\eeq
To the order $g^{2L}$ the Bethe roots $u_{4,j}$
can be simply found from \eq{bae4} \cite{luscherref}.
We will see that explicitly for the single magnon case.

Notice that in the $su(2)$ sector at weak coupling the expression for $Y_{a,0}$ is a rational function
with poles at $u = \pm i\frac{a}{2}$ and $u=u_{4,j}\pm i\frac{a\pm 1}{2}$. As such, for any particular value of $L$ it is straightforward to evaluate the integral in \eq{integ}. For arbitrary $L$ the integrand can be decomposed as
\beqa
\label{ydec}
\frac{Y_{a,0}(u)}{g^{2L}}&\simeq& A(\{u_j\},a,q)\(\frac{4}{a^2+4u^2}\)^L\\
\nn&+&
\sum_{j=1}^{K_4}\sum_{\eta_1=\pm 1,\eta_2=\pm 1}B_{j,\eta_1,\eta_2}(\{u_j\},a,q)
\(\frac{4}{a^2+4u^2}\)^L\frac{1}{u-u_j+i\frac{\eta_1 a+\eta_2}{2}}.
\eeqa
and the integration is done with the use of identities
\beqa
\label{int1}
\int \(\frac{4}{a^2+4u^2}\)^Ldu&=&\sqrt{\pi}\frac{2^{2L-1}\Gamma(L-\tfrac{1}{2})}{a^{2L-1}\Gamma(L)}\\
\int \(\frac{4}{a^2+4u^2}\)^L\frac{1}{2\pi i}\frac{1}{u-i v}du&=&
(-1)^{L+1}\(\frac{2}{a}\)^{2L-1}\frac{\sqrt{\pi}}{2v\Gamma(L)}
{}_2\!{\tilde F}_1\[\frac{1}{2},1;\frac{3}{2}-L;\frac{a^2}{4v^2}\]\nn\\
\label{int2}
&+&\frac{1}{2}\(\frac{4}{a^2-4v^2}\)^L
\eeqa
where $\RE v>0$ and $L$ is large enough\footnote{About ${}_2\!{\tilde F}_1$
see \cite{F12}.}. It would be interesting to see whether expressions similar to \eq{int1}, \eq{int2} come from diagrammatic computations as well.

\subsection{Konishi operator}
As a first application of our method, we will reproduce the results obtained
in \cite{Ahn:2010yv} for the wrapping correction to the $su(2)$ Konishi operator dimension. The $Y_{a,0}$ functions are obtained from \eq{Yaasymp} in terms of the two Bethe roots $u_{4,1},u_{4,2}$, which can be found from the ABA equation \eq{bae4}. At order $g^0$ they are given by
\beqa
	u_{4,1}&=&\frac{(1-3 \Delta)^2}{2 \sqrt{9 \Delta^2-1} \left(3
   \sqrt{1-\Delta^2}+2 \sqrt{\frac{2}{3
   \Delta+1}+1}\right)},\\
   u_{4,2}&=&\frac{(1-3 \Delta)^2}{2 \sqrt{9 \Delta^2-1} \left(3
   \sqrt{1-\Delta^2}-2 \sqrt{\frac{2}{3
   \Delta+1}+1}\right)}
\eeqa
where $\Delta=\frac{\sqrt{5+4\cos\(4\pi\beta\)}}{3}$.
For the Konishi operator $L=4,K_4 = 2$, and hence we have $\tau_{1,\rm L}=\tau_{1,\rm R}=q^4$. Using the expansions \eq{weak1}, \eq{weak2} and the formula \eq{Ta} for the $T_{a,1}$ functions we get from \eq{integ} the following result:
\beqa
	\delta E=-\sum_{a=1}^{\infty}\int\limits \frac{d u }{\pi}  &&\left[
	\(\frac{4g^2}{a^2+4u^2}\)^4
	\frac{Q_4^{[+a-1]}}{Q_4^{[+a+1]}}\frac{Q_4^{[-a+1]}}{Q_4^{[-a-1]}}\times\right.
	\\ \nn
	&& \left.
	\((a+1)-a q^4 \frac{Q_4^{[+a+1]}}{Q_4^{[+a-1]}}
	-a\frac{1}{q^4}\frac{Q_4^{[-a-1]}}{Q_4^{[-a+1]}}
	+(a-1)\frac{Q_4^{[+a+1]}}{Q_4^{[+a-1]}}\frac{Q_4^{[-a-1]}}{Q_4^{[-a+1]}}\)^2\right].
\eeqa
This expression coincides with the wrapping correction given by Eq. (25) of \cite{Ahn:2010yv}. As such, the leading wrapping correction we get is exactly the same as the one obtained in that work. This is an important check of our twisted asymptotic solution of the Y-system.

\subsection{Single magnon momentum quantization}\la{sec:SMS}
Consider now the single magnon case. It is relatively easy to obtain the momentum of a single magnon, as it coincides with the total momentum of the state. It is natural to assume that the total momentum, similarly to the total energy, can be written as
\beq
P=\sum_{j=1}^{K_4}\frac{1}{i}\log\frac{x_{4,j}^+}{x_{4,j}^-}+\delta P\;,
\eeq
with $\delta P$ given by an expression analogous to \eq{Efull}
\beq
\delta P=\sum_{a=1}^\infty\int\frac{du}{2\pi i}\frac{\d p_a^\mir}{\d u}\log(1+Y_{a,0})\;,
\eeq
and the momentum quantization condition then reads
\beq
\sum_{j=1}^{K_4}\frac{1}{i}\log\frac{x_{4,j}^+}{x_{4,j}^-}+\delta P=2\pi \beta + 2 \pi m, \
m \in {\mathbb Z}.
\eeq
This gives
\beq\la{eq:E1m}
E=\sqrt{1+16g^2\sin^2\(\frac{2\pi\beta-\delta P}{2}\)}+\delta E\;,
\eeq
and the {\it exact} position of the Bethe root is given by
\beq
u_{4,1}= \frac{1}{2}\cot\(\frac{2\pi\beta-\delta P}{2}\)\sqrt{1+16g^2\sin^2\(\frac{2\pi\beta-\delta P}{2}\)}\;.
\eeq
Using the expressions for $Y_{a,0}$ from Sec.3
it is straightforward to compute the anomalous dimension of single impurity operators
up to the order $g^{4L-2}$. However, at the moment the perturbation theory results
 are not available beyond order $g^{2L}$.
In the next section we will compute the anomalous dimension to that order
and give an explicit expression for arbitrary $L$ and $\beta$.

\subsection{Single magnon energy at $g^{2L}$ order}\la{sec:1magn}
In this section we compute the energy of a single excitation at the
order $g^{2L}$ for arbitrary $L$ and $\beta$.
For that we notice that in \eq{eq:E1m} the quantity $\delta P$, being of the
order $g^{2L}$, contributes only to the energy at $g^{2L+2}$,
as usual \cite{luscherref}. Thus we can set $u_4 = \half \cot{\pi\beta}$ which is the value of $u_4$ at zeroth order in $g$. We use the weak coupling expansion \eq{weak1}, \eq{weak2}, and the integral in \eq{integ} is straightforward to evaluate, as the integrand is a rational function of $u$. Decomposing $Y_{a,0} (u)$ as in \eq{ydec} and using \eq{int1}, \eq{int2} to integrate the rational functions, we find that the integral equals
\beqa
\label{generI}
{\cal I}(L,a)&=&\sqrt{\pi}(-4)^{L}\frac{\(q-\bar q\)^2\(q^{L-1}-\bar q^{L-1}\)}{\Gamma(L) a^{2L-2}}
\(G_a(q)-G_{-a}(q)-G_a(\bar q)+G_{-a}(\bar q)+aG^0(q)\)\;,\nn\\
&&
\eeqa
where $\bar q=1/q$ and
\beqa
G_a(q)&=&\bar q^{L-1}\frac{a-1}{a(q^2-1)+2}\;\;{}_2\!{\tilde F}_1\[\frac{1}{2},1;\frac{3}{2}-L;\(1+\frac{2}{a(q^2-1)}\)^{-2}\]\;,\\
G^0(q)&=&\frac{(-1)^{L-1}}{2\pi}\(q^{L-1}-\bar q^{L-1}\)\Gamma(L-1/2)\;.
\eeqa
The wrapping correction is given by
$\delta E=\sum_{a=1}^{\infty}{\cal I}(L,a)\;$. We found\footnote{we have checked this explicitly for $L \leq 11$} that instead of summing over $a$ one can equivalently expand the above function
 at $a=0$
\beq
\label{IandA}
{\cal I}(L,a)=\frac{A_{2L-3}(q)}{a^{2L-3}}+\frac{A_{2L-5}(q)}{a^{2L-5}}+\dots\;,
\eeq
and the coefficients in this series expansion give the coefficients in front
of zeta functions in the final result
\beq
\label{finaldE}
\delta E=\sum_{n=L-1}^{2L-3}A_{n}(q)\zeta(n)\;.
\eeq
For fixed $L$ one needs to compute a finite number of terms in the expansion \eq{IandA} of the generating function.

The above result agrees with the perturbation theory calculation. The expression for $\delta E$  obtained with the use of diagrammatic techniques has the following structure\footnote{note that $\lambda$ in \cite{FSSZ,Fiamberti:2010fw} is denoted by $g^2$ in the present paper} \cite{FSSZ1,Fiamberti:2010fw}:
\beq
\label{dEfromdiag}
\delta E=-2L(4 \pi g)^{2L}
\Bigg[(C^{(L)}_{0}-C^{(L)}_{L-1})P^{(L)}
-2\!\sum_{j=0}^{[\frac{L}{2}]-1}\!(C^{(L)}_{j}-C^{(L)}_{L-j-1})I_{j+1}^{(L)}\Bigg]\;,
\eeq
where
\beq
C^{(L)}_{j}=(q-\bar q)^2(q^{2L-2j-2}+\bar q^{2L-2j-2})\;,
\eeq
and $P^{(L)}$ is some known function of $L$, while $I_{j+1}^{(L)}$
represent some particular $L$-loop momentum integrals\footnote{more precisely, their singular parts.
What we denote by $I_{j+1}^{(L)}$ is $\lim_{\varepsilon\to 0}\varepsilon
I_{j+1}^{(L)}(\varepsilon)$ in the original notations.}.
Those integrals were computed in \cite{FSSZ1,Fiamberti:2010fw} explicitly
up to $11$ loops (i.e. for $L \leq 11$), and inserting them into \eq{dEfromdiag} we find complete agreement with our calculations based on \eq{finaldE}.

In fact, those integrals can be directly obtained for any $L$ with the use of  \eq{generI}, \eq{IandA} and \eq{finaldE}. Namely, by inspecting the expression \eq{dEfromdiag}
we notice that the formal expansion about $q=0$ has the following structure modulo some explicit functions
of $L$
\beqa
\label{dEserq}
\delta E&=&\frac{  I^{(L)}_1}{q^{2L}}+
\frac{  I^{(L)}_1+  I^{(L)}_2 }{q^{2L-2}}+
\frac{  I^{(L)}_1+  I^{(L)}_2+  I^{(L)}_3 }{q^{2L-4}}
+
\frac{  I^{(L)}_2+  I^{(L)}_3+  I^{(L)}_4 }{q^{2L-6}}+\dots\;.
\eeqa
We see that by matching the various powers of $q$
with the explicit expressions \eq{finaldE}, \eq{IandA} and \eq{generI} we obtained above one can easily find each of those basis
momentum integrals.

Thus we see that the presence of the deformation parameter $\beta$ (or, equivalently, $q$) allows to extract the perturbation theory integrals $I_{j}^{(L)}$ directly from \eq{dEserq}.

 It would be interesting to repeat this computation in the next to critical
 order in $g$ where the integrals arising and the structure of the result
 in the perturbation theory should be considerable more complicated.
 At the same time for single magnon the Y-system calculation should be possible to do
 up to order $g^{4L-2}$.

\section{Conclusions}

Using the Y-system techniques, we found a general expression for an arbitrary length first wrapping correction for single impurity operators. Our result in the form of a generating function allows to extract directly the relevant Feynman integrals $I_j^{(L)}$, which can be used in the
perturbative calculations. 
In addition, the asymptotic Bethe equations we got in our approach are in complete agreement with the ABA of \cite{BeiRoi}.

We also hope that our results could shed some light on the relation between perturbative
techniques and the AdS/CFT Y-system. It seems that the additional parameter $\beta$ of the deformed theory
could make more transparent the relation and could finally lead to
a derivation of the Y-system directly from perturbation theory. In addition, it would be interesting to investigate more general deformations of ${\cal N}=4$ SYM, and see whether integrability techniques give results consistent with perturbative calculations.

\section*{Acknowledgements}
N.G. thanks P. Vieira for numerous discussions and for collaboration on early stages of this work. We are also grateful to R. Roiban, A. Tseytlin and Z. Tsuboi for useful discussions. The work of NG was partly supported by RFBR project grant RFBR-06-02-16786 and by grant RSGSS-3628.2008.2.
The work of F. L.-M. was partially supported by the
Dynasty Foundation (Russia) and by the grant NS-5525.2010.2.

\section{Appendix A: twisted Bethe equations}
The twisted Bethe equations corresponding to deformations of the ${\cal N}=4$ SYM theory were proposed in \cite{BeiRoi}\footnote{We thank R. Roiban for helpful comments concering the ABA of \cite{BeiRoi}}. In this Appendix we show that under a certain choice of our twists $\tau_{1,{\rm L}}, \tau_{2,{\rm L}}, \tau_{1,{\rm R}}, \tau_{2,{\rm R}}$ those equations coincide\footnote{up to factors of the form $\sigma(u,v)$, which were not known at the time when \cite{BeiRoi} appeared} with Eqs. \eq{ABA}, \eq{bae4fromW} obtained in the Y-system framework. This is true for a general deformation considered in \cite{BeiRoi}, which includes the $\beta$-deformation as a special case.

The general deformation is described in \cite{BeiRoi} by three real parameters $\delta_1, \delta_2,\delta_3$, and the ABA are given by Eq. (5.39) in that work. The notation used in \cite{BeiRoi} is slightly different from ours: $g_{\cite{BeiRoi}} = \sqrt{2} g$, $x_{\cite{BeiRoi}} = g x$.
%, where the quantities with subscript $_{\cite{BeiRoi}}$ are the ones used in \cite{BeiRoi}.
The phases $\delta_i$ enter the ABA through the matrix ${\bf A}$, which is given by Eq. (5.24) in \cite{BeiRoi}.
In our notation, the ABA equations of \cite{BeiRoi} can be written as
\beq
	\left.
	\lambda_1
	\frac{B^{(-)}}{B^{(+)}}
	\frac{Q_2^+}{Q_2^-}
	\right|_{u = u_{1,k}}= 1
,\
	\left.
	\frac{Q_1^+}{Q_1^-} \frac{Q_2^{--}}{Q_2^{++}} \frac{Q_3^+}{Q_3^-}
	\right|_{u = u_{2,k}}= -1
,\
	\left.
	\lambda_1
	\frac{R^{(-)}}{R^{(+)}} \frac{Q_2^+}{Q_2^-}
	\right|_{u = u_{3,k}}= 1\;,
\eeq
\beq
	\left.
	\lambda_2
	\frac{R^{(-)}}{R^{(+)}} \frac{Q_6^+}{Q_6^-}
	\right|_{u = u_{5,k}}= 1
,\
	\left.
	\frac{Q_5^+}{Q_5^-} \frac{Q_6^{--}}{Q_6^{++}} \frac{Q_7^+}{Q_7^-}
	\right|_{u = u_{6,k}}= -1
,\
	\left.
	\lambda_2
	\frac{B^{(-)}}{B^{(+)}}
	\frac{Q_6^+}{Q_6^-}
	\right|_{u = u_{7,k}}= 1\;,
\eeq
\beq
	\left.
	\frac{1}{\lambda_1 \lambda_2}
	\(\frac{x^-}{x^+}\)^L
	\frac{B_1^-}{B_1^+}\frac{B_7^-}{B_7^+}\frac{B_3^+}{B_3^-}\frac{B_5^+}{B_5^-}
	\frac{Q_3^-}{Q_3^+}	\frac{Q_5^-}{Q_5^+} 	\frac{Q_4^{++}}{Q_4^{--}}
	\right|_{u = u_{4,k}}= - 1,
\eeq
where
\beqa
	\lambda_1 &=& 	e^{i(K_4 (\delta_1+\delta_2+\delta_3)+K_5 (-\delta_1-\delta_2-\delta_3)+K_7 		(-\delta_1-\delta_2)+K_1 \delta_3 -L \delta_3 )}\;,\\
	\lambda_2 &=& 	e^{i(K_3 (\delta_1+\delta_2
	+ \delta_3 )+K_4 (- \delta_1-\delta_2-\delta_3)
	-K_7 \delta_1 + K_1 (\delta_2+\delta_3) +L \delta_1)}\;,
\eeqa
and we have used the twisted zero momentum condition (Eq. (5.39) for $j=0$ in \cite{BeiRoi})
\beq
\label{tzemom}
	\prod_{j=1}^{K_4} \frac{x_{4,j}^+}{x_{4,j}^-} = e^{-i(K_4 (\delta_1-\delta_3)-\delta_1 K_5-\delta_1 K_7+\delta_3 K_1+\delta_3 K_3)}
\eeq
which follows from the ABA equations.
We see that those equations coincide with our Eqs. \eq{ABA}, \eq{bae4fromW} if
\footnote{Another possibility is to choose $\tau_{1,{\rm R}}=-\frac{1}{\lambda_1}, \tau_{2,{\rm R}}=-1, \tau_{1,{\rm L}} = -\frac{1}{\lambda_2}, \tau_{2,{\rm L}}=-1$. This gives T-functions which differ by gauge transformation (see \cite{GKV}) from the ones obtained with the choice  \eq{tauvialam1}, and thus the Y-functions and the energy spectrum do not change.
}
\beq
\label{tauvialam1}
	\tau_{1,{\rm R}}=\frac{1}{\lambda_1}, \tau_{2,{\rm R}}=1, 
	\tau_{1,{\rm L}} = \frac{1}{\lambda_2}, \tau_{2,{\rm L}}=1.
\eeq
The $\beta$-deformation, discussed throughout this paper, corresponds \cite{BeiRoi} to the choice
\beq
\la{deltabeta}
	\delta_1 = -2 \pi \beta, \delta_2=0, \delta_3=0.
\eeq
In this case we have
\beq
\label{tau}
	\tau_{1,{\rm R}} = e^{(2K_4-2K_5-2K_7)\pi i\beta},
	\tau_{1,{\rm L}} = e^{(2K_3-2K_4-2K_7+2L)\pi i\beta},
\eeq
and the level matching condition \eq{tzemom} takes the form
\beq
	\prod_{j=1}^{K_4}\frac{x_{4,j}^+}{x_{4,j}^-} = \tau_{1,{\rm R}}.
\eeq
For the $su(2)$ subsector the twists \eq{tau} are in agreement\footnote{Note also that in the $su(2)$ sector exchanging the values of $\tau_{1,\rm L}$ and $\tau_{1,\rm R}$ does not alter the $Y_{a,0}$ functions and the leading wrapping corrections to the anomalous dimensions.} with \eq{twistsu2}.

Note that, as expected (see \cite{BeiRoi}), the twists \eq{tau} are invariant under each of the transformations
\beq
\label{Ks1}
	K_3\to K_3-1,\;K_1\to K_1+1,\;L\to L+1\;
\eeq
\beq
\la{Ks2}
	K_5\to K_5-1,\;K_7\to K_7+1,\;L\to L+1\;.
\eeq

\subsection{Duality transformation and $sl_2$ sector}

The above equations allow to describe an arbitrary state of the $\beta$-deformed theory. However for some applications it may be more convenient to pass to a dualized system of Bethe roots.\footnote
{This section was added after the appearance of \cite{ArMir}, \cite{BomDrinf} to make easier the comparison with results of these papers}
The transformation properties of the transfer matrices under the duality were discussed in \cite{GKV,GFcp3}. To switch from $su(2)$ to $sl(2)$ grading we apply the fermionic duality following \cite{baedual}, along the lines of \cite{GV1loop} and appendix B of \cite{GFcp3}. From Bethe roots $u_{1,j}, \ u_{3,j},\ u_{5,j}, \ u_{7,j}$ we switch to new ones $u_{\tilde1,j}, \ u_{\tilde3,j},\ u_{\tilde5,j}, \ u_{\tilde7,j}$. We first consider the general three-parameter deformation (see previous section). Following \cite{GV1loop} we take
\beq
\la{Kt}
	K_{\tilde 1} = K_2-K_1,\  K_{\tilde 3} = K_2+K_4-K_3,\ 
	K_{\tilde 5} = K_6+K_4-K_5, \ K_{\tilde 7} = K_6-K_7.
\eeq
The new Bethe roots $u_{\tilde1,j}, \ u_{\tilde3,j}$ are related to the original ones by duality relations:
\beq
\label{dual1}
	\frac{R^+_3B^+_1R^+_{\tilde 3}B^+_{\tilde
	1}}{R^-_3B^-_1R^-_{\tilde 3}B^-_{\tilde 1}}
	= 
	\frac{R^{(-)+} Q_2^{++}-\tau_{1,{\rm R}} R^{(+)+} Q_2}{R^{(-)-} Q_2-\tau_{1,{\rm R}} R^{(+)-} Q_2^{--}}
\eeq
\beq
\la{dual2}
	\frac{B^+_3R^+_1B^+_{\tilde 3}R^+_{\tilde
	1}}{B^-_3R^-_1B^-_{\tilde 3}R^-_{\tilde 1}}
	=
	\frac{B^{(-)+} Q_2^{++}-\tau_{1,{\rm R}} B^{(+)+} Q_2}{B^{(-)-} Q_2-\tau_{1,{\rm R}} B^{(+)-} Q_2^{--}}.
\eeq
The corresponding relations for the other (left) wing, which involves $u_5,u_6,u_7$ roots, are obtained by replacing subscripts $\{1,2,3\}\to\{7,6,5\}$ and replacing $\tau_{1,2\;{\rm R}}\to\tau_{1,2\;{\rm L}}$. This remark holds true for all relations in this section.

Using \eq{dual1},\eq{dual2} we can rewrite for example $T_{1,1}$ obtained from \eq{genWR}, \eq{WTsa} as
\beqa
	\nn T_{1,1} &=&
	\frac{1}{\tau_{1,{\rm R}}}\frac{B^{(-)-}}{B^{(+)-}}\frac{Q_1^+}{Q_1^-}
	-\frac{Q_1^+}{Q_1^-}\frac{Q_2^{--}}{Q_2}
	-\frac{Q_3^-}{Q_3^+}\frac{Q_2^{++}}{Q_2}
	+\tau_{1,{\rm R}}\frac{R^{(+)+}}{R^{(-)+}}\frac{Q_3^-}{Q_3^+}
\\
	&=&\frac{1}{f}\(
	-\tilde\tau_{1,{\rm R}}\frac{Q_{\tilde 1}^-}{Q_{\tilde 1}^+}\frac{B^{(+)+}}{B^{(-)+}}
	 +\tilde\tau_{2,{\rm R}}\frac{Q_{\tilde 1}^-}{Q_{\tilde 1}^+}\frac{Q_2^{++}}{Q_2}
	+\frac{1}{\tilde\tau_{2,{\rm R}}}\frac{Q_{\tilde 3}^+}{Q_{\tilde 3}^-}\frac{Q_2^{--}}{Q_2}-
	\frac{1}{\tilde\tau_{1,{\rm R}}}\frac{Q_{\tilde 3}^+}{Q_{\tilde 3}^-}\frac{R^{(-)-}}{R^{(+)-}}
	\)
\eeqa
where the gauge factor is
\beq
	f(u)=
	\frac{1}{\tilde\tau_{1,{\rm R}}}\frac{B_1^-B_{\tilde 1}^-B_3^+B_{\tilde 3}^+}{B_1^+B_{\tilde 1}^+B_3^-B_{\tilde 3}^-}
	\frac{R^{(-)+}}{R^{(+)-}}
\eeq
%with
%\beqa
%\label{taut1}
%	\lambda&=&
%		e^{i(\delta_3(K_{\tilde 1}-2K_2+K_{\tilde 3})-\delta_1(K_{\tilde 5}-2K_6+K_{\tilde 7}))}
%\eeqa
and the twists are
\beqa
	\nn\tilde\tau_{1,{\rm R}}&=&\tilde\tau_{1,{\rm L}}=
		e^{i\half(\delta_3(K_{\tilde 1}-2K_2+K_{\tilde 3})-\delta_1(K_{\tilde 5}-2K_6+K_{\tilde 7}))}\\
	\la{taut1}
	\tilde\tau_{2,{\rm R}}
		&=&e^{i\half((\delta_1+2\delta_2)(K_{\tilde 5}-2K_6+K_{\tilde 7})+
		\delta_3(-K_{\tilde 1}+K_{\tilde 3}+2K_{\tilde 5}-2K_6-2L))}\\
\nn	\tilde\tau_{2,{\rm L}}&=&
		e^{i\half(\delta_1(2K_2-2K_{\tilde 3}-K_{\tilde 5}+K_{\tilde 7}+2L)
		-2\delta_2(K_{\tilde 1}-2K_2+K_{\tilde 3})-\delta_3(K_{\tilde 1}-2K_2+K_{\tilde 3}))}. \;
\eeqa
From \eq{deltabeta} we see that for the $\beta$-deformed theory the twists are given by the above expressions with $\delta_1 = -2 \pi \beta, \delta_2=0, \delta_3=0$.

Note that the transformations \eq{Ks1}, \eq{Ks2} (see \cite{BeiRoi}) do not affect the twists \eq{taut1}, as from \eq{Kt} we see that these transformations amount to
\beq
	K_{\tilde 3}\to K_{\tilde 3}+1,\;K_{\tilde 1}\to K_{\tilde 1}-1,\;L\to L+1\;,
\eeq
\beq
	K_{\tilde 5}\to K_{\tilde 5}+1,\;K_{\tilde 7}\to K_{\tilde 7}-1,\;L\to L+1\;.
\eeq

The duality transformation can also be done on the level of the generating functional. One should use\footnote{it is straightforward to check this e.g. in \it{Mathematica}} (alternatively to \eq{genWR} and \eq{WTsa}) the following functional
\beqa
	\nn\mathcal{W}_{sl(2)} &=&
	 \(1-D\tilde\tau_{1,{\rm R}}\frac{B^{(+)+}Q_{\tilde 1}^-}{B^{(-)+}Q_{\tilde 1}^+}D\)
	 \(1-D\tilde\tau_{2,{\rm R}}\frac{Q_2^{++}Q_{\tilde 1}^-}{Q_2Q_{\tilde 1}^+}D\)^{-1}\\
	 &&
\label{Wsl2}
	 \(1-D\frac{1}{\tilde\tau_{2,{\rm R}}}\frac{Q_2^{--}Q_{\tilde 3}^+}{Q_2Q_{\tilde 3}^-}D\)^{-1}
	 \(1-D\frac{1}{\tilde\tau_{1,{\rm R}}}\frac{Q_{\tilde 3}^+}{Q_{\tilde 3}^-}\frac{R^{(-)-}}{R^{(+)-}}D\)
\eeqa
%\beqa
%	\nn\mathcal{W}_{sl(2)} &=&
%	 \(1-D(\tau_R)^{\half}\frac{B^{(+)+}Q_{\tilde 1}^-}{B^{(-)+}Q_{\tilde 1}^+}D\)
%	 \(1-D\frac{(\tau_R)^{\half}}{\tau}\frac{Q_2^{++}Q_{\tilde 1}^-}{Q_2Q_{\tilde 1}^+}D\)^{-1}\\
%	 &&
%	 \(1-D\frac{\tau}{(\tau_R)^{\half}}\frac{Q_2^{--}Q_{\tilde 3}^+}{Q_2Q_{\tilde 3}^-}D\)^{-1}
%	 \(1-D\frac{1}{(\tau_R)^{\half}}\frac{Q_{\tilde 3}^+}{Q_{\tilde 3}^-}\frac{R^{(-)-}}{R^{(+)-}}D\)
%\eeqa
which is built using the new roots $u_{\tilde1,j}, \ u_{\tilde3,j}$.
The 
$T_{a,s}$ functions are obtained from
\beq
	{{\cal W}}_{sl(2)}=\sum_{s=0}^{\infty}D^s\[ T_{1,s}(u)
	f_s(u)\]D^s\;\;,\;\;
	{{\cal W}}_{sl(2)}^{-1}=\sum_{a=0}^{\infty}(-1)^aD^a\[ T_{a,1}(u)
	f_a(u)\]D^a
\eeq
where the factor
\beq
	f_n(u)=\prod^{\frac{n-1}{2}}_{k=-\frac{n-1}{2}}f(u+ik)
\eeq
corresponds to a gauge transformation on the T-functions \cite{GKV}.
This functional gives BAEs in sl(2) grading as condition of pole cancellation in $T_{1,1}$:
\beq
\label{sl2bae}
	1=\left.\frac{\tilde\tau_{2,{\rm R}}}{\tilde\tau_{1,{\rm R}}}\frac{B^{(-)}}{B^{(+)}}
	\frac{Q^+_2}{Q_2^{-}}\right|_{u=u_{\tilde1,k}}
	,\;\;
	-1=\left.\frac{1}{\(\tilde\tau_{2,{\rm R}}\)^2}
	\frac{Q_{\tilde1}^+Q_2^{--}Q_{\tilde3}^+}{Q_{\tilde1}^-Q_2^{++}Q_{\tilde3}^-}
	\right|_{u=u_{2,k}}
	,\;\;
	1=\left.\frac{\tilde\tau_{2,{\rm R}}}{\tilde\tau_{1,{\rm R}}}\frac{R^{(-)}}{R^{(+)}}
	\frac{Q^+_2}{Q_2^{-}}\right|_{u=u_{\tilde3,k}}\;.
\eeq
These Bethe equations coincide with ones given in Appendix of \cite{BomDrinf}, which can be shown taking into account that the quantity $J$ in that work can be written in our notation as
\beq
	J = \half\(K_{\tilde 1} - K_{\tilde 3} - K_{\tilde 5} + K_{\tilde 7} + 2 L\)
\eeq
(in accordance with (E.22) in \cite{BomDrinf}).

Using \eq{Wsl2} we can establish a relation between T-functions in different gradings. Let us denote by $T_{a,s}^{su(2)}(u|\{u_{1,j}\},\{u_{3,j}\},\tau_{1,{\rm R}},\tau_{2,{\rm R}})$ the T-functions obtained via \eq{WTsa} from the initial functional \eq{genWR}. Denote also by
$T_{a,s}^{sl(2)}(u|\{\tilde u_{1,j}\},\{\tilde u_{3,j}\},\tilde\tau_{1,{\rm R}},\tilde\tau_{2,{\rm R}})$ the T-functions in the $sl(2)$ grading, i.e. the T-functions obtained from $T_{a,s}^{su(2)}(u|\{u_{1,j}\},\{u_{3,j}\},\tau_{1,{\rm R}},\tau_{2,{\rm R}})$ by switching to new Bethe roots $\tilde u_{1,j}, \tilde u_{3,j}$ via duality relations. This means that
\beqa
	T_{a,s}^{sl(2)}(u|\{\tilde u_{1,j}\},\{\tilde u_{3,j}\},\tilde\tau_{1,{\rm R}},\tilde\tau_{2,{\rm R}})=
	T_{a,s}^{su(2)}(u|\{u_{1,j}\},\{u_{3,j}\},\tau_{1,{\rm R}},\tau_{2,{\rm R}})
\eeqa
with the relation between $u_j$ roots and their tilded counterparts being defined by \eq{dual1}, \eq{dual2}. However, $T_{a,s}^{su(2)}$ and $T_{a,s}^{sl(2)}$ have different functional form when we consider their arguments as arbitrary parameters (e.g. for $T_{a,s}^{su(2)}$ these arguments are $\{u_{1,j}\},\{u_{3,j}\},\tau_{1,{\rm R}}$ and $\tau_{2,{\rm R}})$. Nevertheless, it turns out that there are functional relations which allow to write $T_{a,s}^{sl(2)}$ in terms of $T_{a,s}^{su(2)}$. Roughly speaking, these relations amount to exchanging $a$ and $s$ and then taking complex conjugation. Their precise form is:
\beqa
\nn
	T^{sl(2)}_{a,1}(u|\{\tilde u_{1,j}\},\{\tilde u_{3,j}\},\tilde\tau_{1,{\rm R}},\tilde\tau_{2,{\rm R}})
	&=&
	(-1)^a\(f_a(u)\)^{-1}
	\overline{T_{1,a}^{su(2)}(u|\{\tilde u_{1,j}\},\{\tilde u_{3,j}\},
										\tilde\tau_{1,{\rm R}},\tilde\tau_{2,{\rm R}})}\\
										\nn
	T^{sl(2)}_{1,s}(u|\{\tilde u_{1,j}\},\{\tilde u_{3,j}\},\tilde\tau_{1,{\rm R}},\tilde\tau_{2,{\rm R}})
	&=&
	(-1)^s\(f_s(u)\)^{-1}
	\overline{T_{s,1}^{su(2)}(u|\{\tilde u_{1,j}\},\{\tilde u_{3,j}\},
										\tilde\tau_{1,{\rm R}},\tilde\tau_{2,{\rm R}})}\\
	\la{Tcon}
\eeqa
%	\right|_{\{\tau_{1,{\rm R}}\to(\tilde\tau_{1,{\rm R}})^{-1},
%				\;\tau_{2,{\rm R}}\to(\tilde\tau_{2,{\rm R}})^{-1},
%				\;u_{1,j}\to\tilde u_{1,j},
%				\;u_{3,j}\to\tilde u_{3,j},\}}
%	\)
%\eeqa
%	\\
%	\nn
%	\left.T_{1,s}(u|\{u_{1,j}\},\{u_{3,j}\})
%	\right|_{\{\tau_{1,{\rm R}}\to(\tilde\tau_{1,{\rm R}})^{-1},
%				\;\tau_{2,{\rm R}}\to(\tilde\tau_{2,{\rm R}})^{-1}\}}
%	&=&
%	(-1)^s\overline{f_s(u)
%	{T_{s,1}(u|\{\tilde u_{1,j}\},\{\tilde u_{3,j}\},\tau_{1,{\rm R}},\tau_{2,{\rm R}})}}\\
%	\la{Tcon}
%\eeqa
%\beqa
%\nn
%	\left.T_{a,1}(u|\{u_{1,j}\},\{u_{3,j}\})
%	\right|_{\{\tau_{1,{\rm R}}=(\tilde\tau_{1,{\rm R}})^{-1},
%				\;\tau_{2,{\rm R}}=(\tilde\tau_{2,{\rm R}})^{-1}\}}
%	&=&
%	(-1)^a\overline
%	{f_a(u){T_{1,a}(u|\{\tilde u_{1,j}\},\{\tilde u_{3,j}\},\tau_{1,{\rm R}},\tau_{2,{\rm R}})}}\\
%	\nn
%	\left.T_{1,s}(u|\{u_{1,j}\},\{u_{3,j}\})
%	\right|_{\{\tau_{1,{\rm R}}=(\tilde\tau_{1,{\rm R}})^{-1},
%				\;\tau_{2,{\rm R}}=(\tilde\tau_{2,{\rm R}})^{-1}\}}
%	&=&
%	(-1)^s\overline{f_s(u)
%	{T_{s,1}(u|\{\tilde u_{1,j}\},\{\tilde u_{3,j}\},\tau_{1,{\rm R}},\tau_{2,{\rm R}})}}\\
%	\la{Tcon}
%\eeqa
where the bar denotes complex conjugation in the ``physical" plane, i.e. the replacement:
$Q_1^{[+a]}\to Q_{1}^{[-a]}$,
$Q_2^{[+a]}\to Q_{2}^{[-a]}$,
$Q_3^{[+a]}\to Q_{3}^{[-a]}$, 
$R^{(\pm)[\pm a]}\to R^{(\mp)[\mp a]}$ , $B^{(\pm)[\pm a]}\to B^{(\mp)[\mp a]}$,
$\tilde\tau_{1,{\rm R}}\to(\tilde\tau_{1,{\rm R}})^{-1}$,
$\tilde\tau_{2,{\rm R}}\to(\tilde\tau_{2,{\rm R}})^{-1}$. The relations \eq{Tcon} follow from the expressions for generating functionals \eq{genWR} and \eq{Wsl2}, after one takes the Hermitian conjugate of \eq{Wsl2}.
For the undeformed theory (i.e. when $\delta_1=\delta_2=\delta_3=0$) relations \eq{Tcon} reproduce\footnote
{if one takes into account the difference in choice of gauge for the T-functions which amounts to factors $f_n(u)$}
those given in \cite{GKV} for the switch between $sl(2)$ and $su(2)$ gradings.

For example, in the $sl(2)$ sector we have $K_{\tilde1}=K_{\tilde3}=K_{\tilde5}=K_{\tilde7}=K_{2}=K_{6}=0$, and from \eq{taut1} we get
\beq
	\tilde\tau_{1,{\rm R}}=\tilde\tau_{1,{\rm L}}=1,\;
	\tilde\tau_{2,{\rm R}}=e^{-iL\delta_3},\;
	\tilde\tau_{2,{\rm L}} = e^{i L\delta_1}. 
\eeq


\begin{thebibliography}{99}



\bibitem{AdS/CFT} J. M. Maldacena, {\it``The large N limit of superconformal
field theories and supergravity''}, Adv. Theor. Math. Phys. {\bf 2}
231 (1998) [{arXiv:hep-th/9711200}];$::\clubsuit::$
S.~S.~Gubser, I.~R.~Klebanov and A.~M.~Polyakov,
{\it``Gauge theory correlators from non-critical string theory,''}
Phys.\ Lett.\ B {\bf 428} (1998) 105, [hep-th/9802109];$::\clubsuit::$
E.~Witten,
{\it``Anti-de Sitter space and holography,''}
Adv.\ Theor.\ Math.\ Phys.\  {\bf 2} (1998) 253, [hep-th/9802150].
$::\clubsuit::$
  A.~Strominger and C.~Vafa,
  {\it``Microscopic Origin of the Bekenstein-Hawking Entropy,''}
  Phys.\ Lett.\  B {\bf 379} (1996) 99
  [arXiv:hep-th/9601029].
  %%CITATION = PHLTA,B379,99;%%



\bibitem{AdSmore}
  J.~Bagger and N.~Lambert,
  \textit{``Modeling multiple M2's,''}
  Phys.\ Rev.\  D {\bf 75} (2007) 045020
  [arXiv:hep-th/0611108].$::\clubsuit::$
  J.~Bagger and N.~Lambert,
  \textit{``Gauge Symmetry and Supersymmetry of Multiple M2-Branes,''}
  Phys.\ Rev.\  D {\bf 77}, 065008 (2008)
  [arXiv:0711.0955 [hep-th]].$::\clubsuit::$
  %%CITATION = PHRVA,D77,065008;%%
  %%CITATION = PHRVA,D75,045020;%%
      O.~Aharony, O.~Bergman, D.~L.~Jafferis and J.~Maldacena,
    \textit{ ``N=6 superconformal Chern-Simons-matter theories, M2-branes and their
      gravity duals''},
      JHEP {\bf 0810}, 091 (2008)
      [arXiv:0806.1218 [hep-th]]
      %%CITATION = JHEPA,0810,091;%%
    $::\clubsuit::$  O.~Aharony, O.~Bergman and D.~L.~Jafferis,
      \textit{ ``Fractional M2-branes''},
      JHEP {\bf 0811}, 043 (2008)
      [arXiv:0807.4924 [hep-th]],
      %\cite{Strominger:1996sh}



\bibitem{zarads3}
A.~Babichenko, B.~Stefanski and K.~Zarembo,
  \textit{``Integrability and the AdS(3)/CFT(2) correspondence''},
  arXiv:0912.1723.

      %%CITATION = JHEPA,0811,043;%%


\bibitem{adsbae}
J.~A.~Minahan and K.~Zarembo,
{\it``The Bethe-ansatz for N = 4 super Yang-Mills,''}
JHEP {\bf 0303} 013 (2003) [hep-th/0212208].
$::\clubsuit::$ 
M.~Staudacher, {\it``The factorized S-matrix of CFT/AdS,''} {\it JHEP}
{\bf 0505}, 054 (2005) [arXiv:hep-th/0412188].
$::\clubsuit::$
N. Beisert, {\it``The $su(2|2)$ dynamic $S$-matrix,''}
[arXiv:hep-th/0511082];
$::\clubsuit::$
  N.~Beisert, B.~Eden and M.~Staudacher,
  {\it``Transcendentality and crossing,''}
  J.\ Stat.\ Mech.\  {\bf 0701} P021 (2007)
  [arXiv:hep-th/0610251].
$::\clubsuit::$
    J.A.~Minahan and K.~Zarembo,
    \textit{``The Bethe ansatz for superconformal Chern-Simons''},
    {\it JHEP} {\bf 0809}, 040 (2008)
    [arXiv:0806.3951]
    $::\clubsuit::$
    B.I.~Zwiebel,
    \textit{``Two-loop Integrability of Planar ${\cal N}=6$
    Superconformal Chern-Simons Theory''},
    [arXiv:0901.0411]
    $::\clubsuit::$ J.A.~Minahan, W.~Schulgin and K.~Zarembo,
    \textit{``Two loop integrability for Chern-Simons theories
    with ${\cal N}=6$ supersymmetry''},
    {\it JHEP} {\bf 0903}, 057 (2009)
    [arXiv:0901.1142]
    $::\clubsuit::$
    N.~Gromov and P.~Vieira,
    \textit{``The all loop ${\rm AdS}_{4}/{\rm CFT}_{3}$ Bethe ansatz''},
    {\it JHEP} {\bf 0901}, 016 (2009)
    [arXiv:0807.0777].
$::\clubsuit::$
    C.~Ahn and R.I.~Nepomechie,
    \textit{``${\cal N}=6$ super Chern-Simons theory $S$-matrix and all-loop Bethe ansatz
    equations''},
    {\it JHEP} {\bf 0809}, 010 (2008)
    [arXiv:0807.1924].

\bibitem{baedual}
N.~Beisert and M.~Staudacher,
  {\it``Long-range PSU(2,2$|$4) Bethe ansaetze for gauge theory and strings,''}
  Nucl.\ Phys.\ B {\bf 727} 1 (2005)
  [hep-th/0504190].

\bibitem{Crossing}
  R.~A.~Janik,
  {\it ``The AdS(5) x S**5 superstring worldsheet S-matrix and crossing symmetry,''}
  Phys.\ Rev.\  D {\bf 73} 086006 (2006)
  [arXiv:hep-th/0603038].



\bibitem{luscherref}
  Z.~Bajnok and R.~A.~Janik,
  {\it ``Four-loop perturbative Konishi from strings and finite size effects for
  multiparticle states,''}   Nucl.\ Phys.\  B {\bf 807} 625 (2009)
  [arXiv:0807.0399]
$::\clubsuit::$
  Z.~Bajnok, R.~A.~Janik and T.~Lukowski,
  {\it ``Four loop twist two, BFKL, wrapping and strings,''}
  Nucl.\ Phys.\  B {\bf 816}, 376 (2009)
  [arXiv:0811.4448 [hep-th]].
  %%CITATION = NUPHA,B816,376;%%



\bibitem{GKV}  N. Gromov, V. Kazakov, and P. Vieira,
{\it ``Integrability for the Full Spectrum of Planar AdS/CFT,''}
Phys. Rev. Lett. {\bf 103} 131601 (2009) [arXiv:hep-th/0901.3753].




\bibitem{tbarefsads5}
  J.~Ambjorn, R.~A.~Janik and C.~Kristjansen,
  \textit{``Wrapping interactions and a new source of corrections to the spin-chain  /
  string duality,''}
  Nucl.\ Phys.\  B {\bf 736} 288 (2006)
  [arXiv:hep-th/0510171].
$::\clubsuit::$
    G.~Arutyunov and S.~Frolov,
   \textit{``String hypothesis for the $AdS_5 \times S^5$ mirror,''}
  JHEP {\bf 0903} (2009) 152
  [arXiv:0901.1417 [hep-th]].
  %%CITATION = JHEPA,0903,152;%%
$::\clubsuit::$
      D.~Bombardelli, D.~Fioravanti and R.~Tateo,
    \textit{``Thermodynamic Bethe Ansatz for planar AdS/CFT: a proposal,''}
      J.\ Phys.\ A  {\bf 42}, 375401 (2009)
      [arXiv:0902.3930]
      %%CITATION = JPAGB,A42,375401;%%
$::\clubsuit::$
  N.~Gromov, V.~Kazakov, A.~Kozak and P.~Vieira,
  {\it``Exact Spectrum of Anomalous Dimensions of Planar N = 4 Supersymmetric
  Yang-Mills Theory: TBA and excited states,''}
  Lett.\ Math.\ Phys.\  {\bf 91} (2010) 265
  [arXiv:0902.4458 [hep-th]].
  %%CITATION = LMPHD,91,265;%%
$::\clubsuit::$
      G.~Arutyunov and S.~Frolov,
    \textit{``Thermodynamic Bethe Ansatz for the ${\rm AdS}_5 \times {\rm S}^5$ Mirror Model,''}
      JHEP {\bf 0905}, 068 (2009)
      [arXiv:0903.0141].
      %%CITATION = JHEPA,0905,068;%%


\bibitem{tbarefscp3}
%\cite{Bombardelli:2009xz}
  D.~Bombardelli, D.~Fioravanti and R.~Tateo,
  {\it ``TBA and Y-system for planar $AdS_4/CFT_3$,''}
  Nucl.\ Phys.\  B {\bf 834}, 543 (2010)
  [arXiv:0912.4715 [hep-th]].
  %%CITATION = NUPHA,B834,543;%%

\bibitem{GFcp3}
  N.~Gromov and F.~Levkovich-Maslyuk,
  {\it ``Y-system, TBA and Quasi-Classical Strings in AdS4 x CP3,''}
  JHEP {\bf 1006} (2010) 088
  [arXiv:0912.4911 [hep-th]].



\bibitem{LeiStr} R. G. Leigh and M. J. Strassler,
{\it``Exactly marginal operators and duality in
four-dimensional $N = 1$ supersymmetric gauge theory,''}
Nucl. Phys. {\bf B447} 95 (1995) [arXiv:hep-th/9503121].


\bibitem{MPSZ} A. Mauri, S. Penati, A. Santambrogio, and D. Zanon,
{\it ``Exact results in planar $N = 1$
superconformal Yang-Mills theory,''} JHEP {\bf 11} 024 (2005) [arXiv:hep-th/0507282].




\bibitem{LunMal} O. Lunin and J. M. Maldacena, {\it ``Deforming field theories with
$U(1)\times U(1)$
global symmetry and their gravity duals,''} JHEP {\bf 05} 033 (2005) [hep-th/0502086].



\bibitem{Frolov} S. Frolov, {\it``Lax pair for strings in Lunin-Maldacena background,''}
JHEP {\bf 05} 069 (2005) [hep-th/0503201].



\bibitem{BykFro} D. V. Bykov and S. Frolov, {\it``Giant magnons in TsT-transformed
$AdS_5\times S^5$,''} JHEP {\bf 07} 071 (2008)  [arXiv:hep-th/0805.1070].

%\cite{Frolov:2005iq}

\bibitem{FrRoTs} S. A. Frolov, R. Roiban, and A. A. Tseytlin,
{\it ``Gauge - string duality for superconformal deformations of $N = 4$ super
Yang-Mills theory,''} JHEP {\bf 07} 045 (2005) [hep-th/0503192].

\bibitem{Frolov:2005iq}
  S.~A.~Frolov, R.~Roiban and A.~A.~Tseytlin,
  {\it ``Gauge-string duality for (non)supersymmetric deformations of N = 4  super
  Yang-Mills theory,''}
  Nucl.\ Phys.\  B {\bf 731} (2005) 1
  [arXiv:hep-th/0507021].
  %%CITATION = NUPHA,B731,1;%%


\bibitem{BerChe} D. Berenstein and S. A. Cherkis, {\it``Deformations of $N = 4$ SYM
and integrable spin chain models,''} Nucl. Phys. {\bf B702} 49 (2004) [hep-th/0405215].


\bibitem{BeiRoi} N. Beisert and R. Roiban, {\it``Beauty and the twist:
The Bethe ansatz for twisted $N = 4$ SYM,''} JHEP {\bf 08} 039 (2005) [hep-th/0505187].



\bibitem{Roiban:2003dw}
  R.~Roiban,
  {\it ``On spin chains and field theories,''}
  JHEP {\bf 0409} (2004) 023
  [arXiv:hep-th/0312218].
  %%CITATION = JHEPA,0409,023;%%


\bibitem{FSSZ} F. Fiamberti, A. Santambrogio, C. Sieg, and D. Zanon,
{\it``Finite-size effects in the superconformal $\beta$-deformed ${\cal N} = 4$ SYM,''}
JHEP {\bf 08} 057 (2008) [hep-th/0806.2103].



\bibitem{FSSZ1} F. Fiamberti, A. Santambrogio, C. Sieg, and D. Zanon,
{\it``Single impurity operators at critical wrapping order in the beta-deformed
${\cal N} = 4$ SYM,''} JHEP {\bf 08} 034 (2009) [hep-th/0811.4594].

%\cite{Fiamberti:2010fw}


\bibitem{Fiamberti:2010fw}
  F.~Fiamberti, A.~Santambrogio and C.~Sieg,
  {\it ``Superspace methods for the computation of wrapping effects in the standard
  and beta-deformed N=4 SYM,''}
  arXiv:1006.3475 [hep-th].
  %%CITATION = ARXIV:1006.3475;%%

%\cite{Gunnesson:2009nn}


\bibitem{Gunnesson:2009nn}
  J.~Gunnesson,
  {\it ``Wrapping in maximally supersymmetric and marginally deformed N=4
  Yang-Mills,''}
  JHEP {\bf 0904} (2009) 130
  [arXiv:0902.1427 [hep-th]].
  %%CITATION = JHEPA,0904,130;%%

%\cite{Beccaria:2009hg}


\bibitem{Beccaria:2009hg}
  M.~Beccaria and G.~F.~De Angelis,
  {\it ``On the wrapping correction to single magnon energy in twisted N=4 SYM,''}
  Int.\ J.\ Mod.\ Phys.\  A {\bf 24} (2009) 5803
  [arXiv:0903.0778 [hep-th]].
  %%CITATION = IMPAE,A24,5803;%%

%\cite{Gromov:2008gj}


\bibitem{Ahn:2010yv}
  C.~Ahn, Z.~Bajnok, D.~Bombardelli and R.~I.~Nepomechie,
  {\it ``Finite-size effect for four-loop Konishi of the beta-deformed N=4 SYM,''}
  arXiv:1006.2209 [hep-th].
  %%CITATION = ARXIV:1006.2209;%%




    \bibitem{Fiamberti:2008sh}
      F.~Fiamberti, A.~Santambrogio, C.~Sieg and D.~Zanon,
      \textit{``Anomalous dimension with wrapping at four loops in N=4 SYM,''}
      Nucl.\ Phys.\  B {\bf 805}, 231 (2008)
      [arXiv:0806.2095]
      %%CITATION = NUPHA,B805,231;%%
    $::\clubsuit::$
    V.~N.~Velizhanin,
      \textit{``Leading transcedentality contributions to the four-loop universal anomalous
      dimension in N=4 SYM,''}
      arXiv:0811.0607.



    \bibitem{Fiamberti:2009jw}
      F.~Fiamberti, A.~Santambrogio and C.~Sieg,
      \textit{``Five-loop anomalous dimension at critical wrapping order in N=4 SYM,''}
      arXiv:0908.0234 .
      %%CITATION = ARXIV:0908.0234;%%

%\cite{Ahn:2010yv}


\bibitem{Zamolodchikov:1991et}
  C.~N.~Yang and C.~P.~Yang,
  \textit{``One-dimensional chain of anisotropic spin-spin interactions. I: Proof of
  Bethe's hypothesis for ground state in a finite system,''}
  Phys.\ Rev.\  {\bf 150} (1966) 321.
  %%CITATION = PHRVA,150,321;%%
$::\clubsuit::$
  A.~B.~Zamolodchikov,
  \textit{``On the thermodynamic Bethe ansatz equations for reflectionless ADE
  scattering theories,''}
  Phys.\ Lett.\  B {\bf 253}, 391 (1991).
  %%CITATION = PHLTA,B253,391;%%
$::\clubsuit::$
%\bibitem{Dorey:2006dq}
  N.~Dorey,
\textit{``Magnon bound states and the AdS/CFT correspondence,''}
  J.\ Phys.\ A  {\bf 39}, 13119 (2006)
  [arXiv:hep-th/0604175].
  %%CITATION = JPAGB,A39,13119;%%
$::\clubsuit::$
%bibitem{Takahashi}
M.~Takahashi, \textit{``Thermodynamics of one-dimensional solvable models"},
Cambridge University Press, 1999.
$::\clubsuit::$
%bibitem{KorepinBook}
F.H.L. Essler, H.Frahm, F.G\"ohmann, A. Kl\"umper and V. Korepin,
\textit{"The One-Dimensional Hubbard Model"},  Cambridge University Press, 2005.
$::\clubsuit::$
%
%bibitem{Bazhanov:1996aq}
  V.~V.~Bazhanov, S.~L.~Lukyanov and A.~B.~Zamolodchikov,
  \textit{``Quantum field theories in finite volume: Excited state energies,''}
  Nucl.\ Phys.\  B {\bf 489}, 487 (1997)
  [arXiv:hep-th/9607099].
  %%CITATION = NUPHA,B489,487;%%
$::\clubsuit::$
%bibitem{Dorey:1996re}
  P.~Dorey and R.~Tateo,
  \textit{``Excited states by analytic continuation of TBA equations,''}
  Nucl.\ Phys.\  B {\bf 482}, 639 (1996)
  [arXiv:hep-th/9607167].
  %%CITATION = NUPHA,B482,639;%%
$::\clubsuit::$
%bibitem{Fioravanti:1996rz}
  D.~Fioravanti, A.~Mariottini, E.~Quattrini and F.~Ravanini,
  \textit{``Excited state Destri-De Vega equation for sine-Gordon and restricted
  sine-Gordon models,''}
  Phys.\ Lett.\  B {\bf 390}, 243 (1997)
  [arXiv:hep-th/9608091].
  %%CITATION = PHLTA,B390,243;%%
$::\clubsuit::$
%bibitem{Bytsko:2006ut}
  A.~G.~Bytsko and J.~Teschner,
\textit{``Quantization of models with non-compact quantum group symmetry: Modular
  XXZ magnet and lattice sinh-Gordon model,''}
  J.\ Phys.\ A  {\bf 39} (2006) 12927
  [arXiv:hep-th/0602093].
  %%CITATION = JPAGB,A39,12927;%%
$::\clubsuit::$
  N.~Gromov, V.~Kazakov and P.~Vieira,
  {\it``Finite Volume Spectrum of 2D Field Theories from Hirota Dynamics,''}
  JHEP {\bf 0912} (2009) 060
  [arXiv:0812.5091 [hep-th]].
  %%CITATION = JHEPA,0912,060;%%
$::\clubsuit::$
%bibitem{Saleur:2009bf}
  H.~Saleur and B.~Pozsgay,
\textit{``Scattering and duality in the 2 dimensional $OSP(2|2)$ Gross Neveu and sigma
  models,''}
  arXiv:0910.0637.
  %%CITATION = ARXIV:0910.0637;%%


\bibitem{Cavaglia:2010nm}
  A.~Cavaglia, D.~Fioravanti and R.~Tateo,
  \textit{``Extended Y-system for the $AdS_5/CFT_4$ correspondence,''}
  arXiv:1005.3016 [hep-th].
  %%CITATION = ARXIV:1005.3016;%%


\bibitem{GKVKonishi}
  N.~Gromov, V.~Kazakov and P.~Vieira,
  {\it``Exact Spectrum of Planar ${\cal N}=4$ Supersymmetric Yang-Mills Theory:
  Konishi Dimension at Any Coupling,''}
  Phys.\ Rev.\ Lett.\  {\bf 104}, 211601 (2010)
  [arXiv:0906.4240 [hep-th]].
  %%CITATION = PRLTA,104,211601;%%



\bibitem{ads5quasicl}
    %\cite{Gromov:2009tq}
      N.~Gromov,
      \textit{``Y-system and Quasi-Classical Strings,''}
      arXiv:0910.3608.
      %%CITATION = ARXIV:0910.3608;%%
    $::\clubsuit::$
  N.~Gromov, V.~Kazakov and Z.~Tsuboi,
  {\it ``PSU$(2,2|4)$ Character of Quasiclassical AdS/CFT,''}
  arXiv:1002.3981 [hep-th].
  %%CITATION = ARXIV:1002.3981;%%



\bibitem{tbanum}
  G.~Arutyunov, S.~Frolov and R.~Suzuki,
  {\it ``Exploring the mirror TBA,''}
  JHEP {\bf 1005} (2010) 031
  [arXiv:0911.2224 [hep-th]].
  %%CITATION = JHEPA,1005,031;%%
    $::\clubsuit::$
  S.~Frolov,
  {\it ``Konishi operator at intermediate coupling,''}
  arXiv:1006.5032 [hep-th].
  %%CITATION = ARXIV:1006.5032;%%



\bibitem{GV1loop}
N.~Gromov and P.~Vieira,
  {\it ``Complete 1-loop test of AdS/CFT,''}
  JHEP {\bf 0804} (2008) 046
  [arXiv:0709.3487 [hep-th]].
  %%CITATION = JHEPA,0804,046;%%
\bibitem{Bazhanov:2010ts}
  V.~V.~Bazhanov, T.~Lukowski, C.~Meneghelli and M.~Staudacher,
  {\it ``A Shortcut to the Q-Operator,''}
  arXiv:1005.3261 [hep-th].
  %%CITATION = ARXIV:1005.3261;%%

%\cite{Kazakov:2007fy}


\bibitem{Tsuboi:1997iq}
 Z.~Tsuboi,
 {\it``Analytic Bethe ansatz and functional equations for Lie superalgebra
 $sl(r+1|s+1)$,''}
 J.\ Phys.\ A  {\bf 30}, 7975 (1997).
 %%CITATION = JPAGB,A30,7975;%%

 \bibitem{Tsuboi:1998ne}
 Z.~Tsuboi,
 {\it``Analytic Bethe Ansatz And Functional Equations Associated With Any Simple
 Root Systems Of The Lie Superalgebra $sl(r+1|s+1)$,''}
 Physica A {\bf 252}, 565 (1998).
 %%CITATION = PHYSA,A252,565;%%


%(2)
%\cite{Tsuboi:1998ne}

\bibitem{Kazakov:2007fy}
  V.~Kazakov, A.~S.~Sorin and A.~Zabrodin,
  {\it``Supersymmetric Bethe ansatz and Baxter equations from discrete Hirota
  dynamics,''}
  Nucl.\ Phys.\  B {\bf 790} (2008) 345
  [arXiv:hep-th/0703147].
  %%CITATION = NUPHA,B790,345;%%

%\cite{Zabrodin:2007rq}


\bibitem{Zengo}
  N.~Gromov, V.~Kazakov, S.~Leurent and Z.~Tsuboi,
{\it ``Wronskian Solution for AdS/CFT Y-system,''} arXiv:1010.2720 [hep-th].

%(1)
%\cite{Tsuboi:1997iq}

\bibitem{Zabrodin:2007rq}
  A.~Zabrodin,
  {\it ``Backlund transformations for difference Hirota equation and supersymmetric
  Bethe ansatz,''}
  arXiv:0705.4006 [hep-th].
  %%CITATION = ARXIV:0705.4006;%%

%\cite{Bazhanov:2010ts}



\bibitem{Serban:2010sr}
  D.~Serban,
  {\it``Integrability and the AdS/CFT correspondence,''}
  arXiv:1003.4214 [hep-th].
  %%CITATION = ARXIV:1003.4214;%%


\bibitem{F12}
http://functions.wolfram.com/HypergeometricFunctions/Hypergeometric2F1Regularized/

\bibitem{ArMir}
  G.~Arutyunov, M.~de Leeuw and S.~J.~van Tongeren,
  {\it ``Twisting the Mirror TBA,''}
  arXiv:1009.4118 [hep-th].
  %%CITATION = ARXIV:1009.4118;%%


\bibitem{BomDrinf}
 C.~Ahn, Z.~Bajnok, D.~Bombardelli and R.~I.~Nepomechie,
  {\it ``Twisted Bethe equations from a twisted S-matrix,''}
  arXiv:1010.3229 [hep-th].
  %%CITATION = ARXIV:1010.3229;%%


    \end{thebibliography}
    \end{document}